\documentclass[epj,final]{svjour}

\usepackage{latexsym}
\usepackage{url}
\usepackage{amsfonts}
\usepackage{amsmath, amssymb}
\RequirePackage{graphicx}

\begin{document}
\title{Lense-Thirring precession and modified gravity constraints}
\author{A.~Stepanian\inst{1}, Sh.~Khlghatyan\inst{1}
}                     
%
%
\institute{Center for Cosmology and Astrophysics, Alikhanian National Laboratory and Yerevan State University, Yerevan, Armenia}
\date{Received: date / Revised version: date}
%

\abstract{The orbital Lense-Thirring precession is considered in the context of constraints for weak-field General Relativity involving the cosmological constant  $\Lambda$. It is shown that according to the current accuracy of satellite measurements the obtained error limits for $\Lambda$ is self-consistent with cosmological observations. The corrections of $\Lambda$ term are derived for the strong field Lense-Thirring precession i.e. the frame dragging effect and for the nutation. As a result, in the context of recently proposed $\Lambda$-gravity we obtain constraints for $\Lambda$ in both relativistic and weak-field limits. Namely, for the latter we analyze several Keplerian systems at different scales. We find that the obtained constraints for the modified gravity corrections  are several orders of magnitude tighter than those available for such effects as gravitational redshift, gravitational time delay and geodetic precession in Solar System. }

%
\maketitle

\section{Introduction}
The Newton theorem on ``sphere-point" equivalency leads to a unified picture for describing the dark sector (DS)\cite{G,GS1,GS2}. Namely, within this approach the cosmological constant $\Lambda$, as a fundamental constant \cite{GS3}, enters naturally in the weak-field limit of General Relativity (GR). 

In this sense, on one hand the presence of $\Lambda$ in GR equations explains the accelerated expansion of the Universe as a candidate for dark energy (DE), on the other hand, it enables one to describe the dynamics of dark matter (DM) in galactic configurations. Consequently, the weak-field modification of GR is given by the following metric \cite{GS1}
\begin {equation} \label {mod}
g_{00} = 1 - \frac{2 G m}{c^2r} - \frac{\Lambda r^2}{3}\,; 
\qquad g_{rr} = \left(1 - \frac{2 G m}{c^2r} - \frac {\Lambda r^2}{3}\right)^{-1}\,.
\end {equation} 
Indeed, the general function for force $\mathbf{F}(r)$ satisfying Newton's theorem used in the above metric has the form (see \cite{G,GS1,G1}) 
\begin{equation}\label{FandU}
\mathbf{F}(r) = \left(-\frac{A}{r^2} + Br\right)\hat{\mathbf{r}}\, .
\end{equation}     
It should be noticed that, although the above metric was known before as Schwarzschild-de Sitter (SdS) metric, this approach enables one to describe the galaxy clusters within the weak-field limit of GR \cite{GS2}. 

Within group-theoretical approach the isometry groups are defined depending on the sign of $\Lambda$ for three different vacuum solutions for GR equations, as given in Table \ref{tab1}.

\begin{table}
\caption{}\label{tab1}
\centering
\begin{tabular}{ |p{1.5cm}||p{3.6cm}|p{3cm}|p{1.8cm}|}
\hline
\multicolumn{4}{|c|}{Background geometries} \\
\hline
Sign& Spacetime&Isometry group&Curvature\\
\hline
$\Lambda > 0$ &de Sitter (dS) &O(1,4)&+\\
$ \Lambda = 0$ & Minkowski (M) & IO(1,3)&0 \\
$\Lambda <0 $ &Anti de Sitter (AdS) &O(2,3)&-\\
\hline
\end{tabular}
\end{table}
\noindent 
The stabilizer group of these maximally symmetric Lorentzian 4D-geometries is the Lorentz group O(1,3). In this sense, for all of these Lorentzian geometries, the group O(1,3) of orthogonal transformations implies a spherical symmetry (in Lorentzian sense) at each point i.e. 
\begin{equation}\label{qspace}
dS= \frac{O(1,4)}{O(1,3)}\,, \quad M=\frac{IO(1,3)}{O(1,3)}\,, \quad AdS=\frac{O(2,3)}{O(1,3)} \,.
\end{equation}
The full Poincare group IO(1,3) is reduced to Galilei group Gal(4)=(O(3)$\times$R)$\ltimes$R${}^{6}$ in the non-relativistic limit, as an action of O(3)$\times$R  on group of boosts and spatial translations R${}^{6}$. Then, for the non-relativistic limit of O(1,4) and O(2,3) groups one has
\begin{equation}\label{nrel}
O(1,4) \to (O(3) \times O(1,1)) \ltimes R^6\,,\quad O(2,3) \to (O(3) \times O(2) ) \ltimes R^6\,.
\end{equation}
At the same time, the Galilei spacetime appears via quotienting Gal(4) by O(3)$\times$R${}^{3}$, while the Newton-Hooke NH${}^{\pm}$ spacetimes are given by the same quotient group but for groups achieved in Eq.(\ref{nrel}); O(3) for these cases is the stabilizer group of spatial geometry which at each point admits O(3) symmetry. 
 
The next important fact is that, the force of Eq.(\ref{FandU}) defines non-force-free field inside a spherical shell, thus drastically contrasting with Newton's gravity when the shell has force-free field in its interior. The non-force-free field agrees with observational indications that galactic halos do determine features of galactic disks \cite{Kr}. The weak-field GR is able to describe the observational features of galactic halos \cite{G,Ge}, of groups and clusters of galaxies \cite{GS2}. Several other problems such as the stability of $N$-body gravitating systems have been studied in the context of $\Lambda$-gravity \cite{GKS}. Meantime, $\Lambda$-gravity proposes a natural solution for the so-called ``H tension" problem \cite{GS4}. Currently, the cosmological observations \cite{Pl} indicate that $\Lambda = 1.11 \times 10^{-52}$ $m^{-2}$. Although, this numerical value is too small to be detected directly via observations, it is expected that in the recent future via more accurate measurements of gravitational lensing, we will be able to mark the discrepancy between the $\Lambda$-gravity and the standard GR theory \cite{GS6}.

In what follows, we study the orbital Lense-Thirring (LT) precession and nutation in the context of $\Lambda$-gravity, as effects regarded as important ones to affect the standard Keplerian orbital motion \cite{Shant}. For both of them, we obtain the constraints of $\Lambda$  and compare them with the current observational data. Namely, this paper can be regarded as the continuation of several analyses in which by considering both relativistic and non-relativistic effects, different upper constraints have been reported for $\Lambda$ \cite{M,J1,J2,L}.

\section{Frame dragging}

Frame dragging is one of the important predictions of GR. It was predicted in 1918 by J. Lense and H. Thirring \cite {LT1,LT2,LT3} as the precession of gyroscope orbiting a rotating body of mass $M$ and angular momentum $J$ at spacetime metric

\begin {equation} \label {LT}
ds^2 = (1-\frac{2 G M}{r c^2})c^2 dt^2 - (1+\frac{2 G M}{r c^2}) d\sigma^2 + 4 G\epsilon _{ijk}J^k \frac{ x^i}{c^3 r^3} c dt dx^j,
\end {equation}
where $d\sigma^2 = dx^2 + dy^2 + dz^2$ is the line element of Euclidean 3-geometry. Thus the rate of precession can be achieved by solving the geodesic equation \cite {AB,LL,MTW,W}
\begin {equation} \label {Pr}
\Omega_{GR} = \frac {2GJ}{c^2\widetilde{a}^3 (1-e^2)^{3/2}},
\end {equation}
where $\widetilde{a}$ and $e$ are the semi-major axis and the eccentricity of the orbit, respectively. 

The rate of procession as in Eq.(\ref{Pr}) is one of the predictions of Einstein's theory and its measurement for the Earth's gravity is considered as one of the accurate tests of GR \cite{LT4,LT5}. By now, the most accurate measurement of this effect was performed by means of LARES satellite, with a result compatible to GR to few percent accuracy \cite{LARES,LARES1}. It is expected that this limit can be improved by forthcoming mission LARES-2 \cite{LARES2I,LARES2II}.

In the context of $\Lambda$-gravity the metric in Eq.(\ref{LT}) will be written as
\begin {equation} \label {LTL}
ds^2 = (1-\frac{2 G M}{r c^2} - \frac {\Lambda r^2}{3})c^2 dt^2 - (1+\frac{2 G M}{r c^2} + \frac {\Lambda r^2}{3}) d\sigma^2 + 4 G\epsilon _{ijk}J^k \frac{ x^i}{c^3 r^3} c dt dx^j.
\end {equation}
Consequently, the rate of precession changes to
\begin {equation} \label{PrL}
\Omega_{tot}=\Omega_{GR}+\Omega_\Lambda = \frac {2GJ}{c^2 \widetilde{a}^3 (1-e^2)^{3/2}} + \frac{\Lambda J}{3 M}.
\end {equation}
It should be noticed that, in the above equation the $\Omega_{\Lambda}$ term has no dependence on the distance of the gyroscope or any other orbital parameters. Particularly, the correction of LT in the context of $\Lambda$-gravity will be an additional term which is related to the mass and angular momentum of central object. From a fundamental point of view, the $\Omega_{\Lambda}$ term shows that in contrast to corrections of $\Lambda$-gravity in other effects, the presence of $\Lambda$ cannot be interpreted as an additional exotic matter with density equal to $\frac{-\Lambda c^2}{4 \pi G}$. Moreover, the presence of $\Lambda$ in Eq.(\ref{PrL}) is due to pure relativistic corrections. Such statement can be verified by checking the couplings of $\Lambda$ term too. Namely, since the LT itself is a pure relativistic effect which cannot be reduced to any classical analogue, the $\Lambda$ term is not coupled to $c^2$ anymore. 

\section{Strong field analysis}

It should be noticed that, LT metric in Eq. (\ref{LT}) can be considered as the weak-field limit for Kerr metric

\begin {equation} \label {Kerr}
ds^2 = \frac {\Delta_r}{\rho^2}(cdt-a\sin^2\theta d\phi)^2 - \frac{\rho^2}{\Delta_r} dr^2 - \rho^2 d\theta^2 - \frac { \sin ^2 \theta}{ \rho^2} (a cdt - (r^2+a^2)d\phi)^2.
\end {equation}
where $\Delta_r= r^2 - \frac{2 G M r}{c^2} +a^2$ and $\rho^2 = r^2+ a^2 \cos ^2 \theta$. The Kerr parameter $a$ is defined as $\frac{J}{M c}$. According to this metric the frame dragging is defined as
\begin {equation} \label {FD}
\Omega = - \frac {g_{t\phi}}{g_{\phi\phi}}.
\end {equation} 
Clearly for $a\ll 1$, the Kerr metric is reduced to LT metric. Thus, in order to verify our results derived in Eq.(\ref{PrL}), it is essential to check them starting from Kerr metric with $\Lambda$, and deriving $\Omega$. In this sense, the original Kerr metric is modified to
\begin {equation} \label {KerrL}
ds^2 = \frac {\Delta_r}{\rho^2 L^2}(cdt-a\sin^2\theta d\phi)^2 - \frac{\rho^2}{\Delta_r} dr^2 - \frac{\rho^2}{\Delta_\theta} d\theta^2 - \frac {\Delta_\theta \sin ^2 \theta}{\rho^2 L ^2} (a cdt - (r^2+a^2)d\phi)^2. 
\end {equation}
with the following parameters
\begin {equation} 
\begin{aligned}
&\Delta_r = (1- \frac{\Lambda r^2}{3})(r^2+a^2) - \frac{2 G M r}{c^2}, \\
&\Delta_\theta = (1+ \frac{a^2 \Lambda \cos^2 \theta}{3}), \\
&L = (1+ \frac{a^2 \Lambda}{3}), \\
&\rho^2 = r^2+ a^2 \cos ^2 \theta.
\end{aligned}
\end {equation}
It should be recalled that, similar to SdS metric, Eq.(\ref{KerrL}) was known before as Kerr-de Sitter metric and was used to describe the Universe when a single axially symmetric object is immersed in de Sitter background \cite{RM}. However, by considering Newton theorem it can be used to obtain the LT metric for every slowly rotating object according to Eq.(\ref{LTL}).

Now, according to Eq.(\ref{FD}), the rate of precession will be
\begin {equation} \label {FDL}
\Omega = - \frac { \frac{a c 2 G M r}{c^2} +\frac{c \Lambda}{3}(a r^4 + a^3 r^2+ a^3 r^2\cos^2 \theta + a^5 cos^2 \theta)}{a^2 \sin^2 \theta (r^2 +a^2- \frac {\Lambda}{3} (r^4 + a^2 r^2 )-\frac{2 G M r}{c^2} )-( r^4 + a^4 + 2 a^2 r^2) (1+ \frac{a^2 \Lambda \cos ^2 \theta}{3})}.
\end {equation} 
For $a \ll 1$ we get
\begin {equation} \label {FDLL}
\Omega = \frac {\frac{a c 2 G M r}{c^2} +\frac{c \Lambda (a r^4)}{3}}{r^4},
\end {equation} 
where after some simple algebra one can recover Eq.(\ref{PrL}). 

It is worth to mention that in Eq.(\ref{FDLL}) the effect of $\Lambda$ becomes important at large distances from the object. The distance beyond which the second term in Eq.(\ref{FDLL}) becomes dominant is
\begin {equation} \label {Imp}
{r^3}= \frac{ 6 G M }{\Lambda c^2}.
\end {equation}
Considering the current value of $\Lambda$ according to \cite{Pl} i.e. $1.11 \times 10^{-52}$ $m^{-2}$, this distance will be located far away from the rotating object. This radius for typical objects is tabulated in Table \ref{tab2}.

\begin{table}
\caption{}\label{tab2}
\centering
\begin{tabular}{ |p{2.4cm}||p{2.7cm}|p{1.8cm}| }
\hline
\multicolumn{3}{|c|}{Critical distance for different objects} \\
\hline
Central Object& Mass (Kg)&Radius (m)\\
\hline
Earth &5.97 $\times 10^{24}$ & 6.21 $\times 10^{16}$  \\
\hline
Sun & 1.98 $\times 10^{30}=M_{\odot}$ & 4.30 $\times 10^{18}$ \\
\hline
Sgr A${}^{*}$  &4.28 $\times 10^6 M_{\odot}$& 6.98 $\times 10^{20}$ \\
\hline
\end{tabular}
\end{table}

\section{Constraints for $\Lambda$}

Considering Eq.(\ref{PrL}), we can obtain the error limits for $\Lambda$. Namely, we have calculated the error limits of $\Lambda$ for LAGEOS-1, LAGEOS-2 , LARES and the forthcoming LARES-2 \cite{LARES2II} based on their reported or predicted  accuracies. The results are shown in Table \ref{tab3}. 
\begin{table}
\caption{}\label{tab3}
\centering
\begin{tabular}{ |p{1.8cm}||p{1.6cm}|p{3cm}|p{1.8cm}| p{2.4cm}|}
\hline
\multicolumn{5}{|c|}{Error limits of $\Lambda$ for LT effect} \\
\hline
Satellite& accuracy & semi major axis $(m)$ & eccentricity  &$\Lambda $ ($m^{-2}$)$<$   \\
\hline
LAGEOS-1 & 0.200 & 12271150.0 & 0.004456 & 	$2.874459 \times 10^{-24}$  \\
\hline
LAGEOS-2 &	0.200 & 12161840.0 & 0.013730 & $2.953411 \times 10^{-24}$ \\
\hline
LARES & 0.050 & 7822000.0 & 0.000800 & 	$2.774508 \times 10 ^{-24}$ \\
\hline
LARES 2 &	0.002 & 12270000.0 & 0.002500 &	$2.875208 \times 10^{-26}$ \\
\hline
\end{tabular}
\end{table}

Here, it should be noticed that in both Eqs.(\ref{Pr},\ref{PrL}) one uses a simplified view of the gravitational field of the Earth; for the contribution of high tidal modes see  \cite{VG}. 
For corrections up to the second order we get the following relation
\begin{equation}\label{Real}
\Lambda \leq \mathbb{E}(\Omega) \frac{3M}{J}(\frac{ G M R^2 \omega}{5 c^2 r^3 })(1 - (\frac{219}{392})J_2(\frac{R}{r})^2),
\end{equation}
where $\mathbb{E}(\Omega)$ is the accuracy of the measurement,  $R$ and $\omega$  are the radius of the Earth and its angular frequency, respectively, $r$ is the distance from the satellite to Earth's center and $J_2$ is the Bessel function of first kind. The upper limits of the error for $\Lambda$ are given in Table \ref{tab4}. In this case, the limits are improved and become tighter. The obtained limits show that the predictions of LT precession in the context of $\Lambda$-gravity are fully consistent with the observational data in \cite{Pl}, as in all cases the numerical value of $\Lambda$ lies in the error limits.
\begin{table}
\caption{}\label{tab4}
\centering
\begin{tabular}{ |p{1.8cm}||p{1.6cm}|p{3cm}|p{1.8cm}| p{2.4cm}|}
\hline
\multicolumn{5}{|c|}{Error limits of $\Lambda$ for LT effect in more realistic case} \\
\hline
Satellite& accuracy & semi major axis $(m)$ & eccentricity  &$\Lambda $ ($m^{-2}$)$<$   \\
\hline
LAGEOS-1 & 0.200 & 12271150.0 & 0.004456 & 	$1.881532 \times 10^{-31}$  \\
\hline
LAGEOS-2 &	0.200 & 12161840.0 & 0.013730 & $1.932717 \times 10^{-31}$ \\
\hline
LARES & 0.050 & 7822000.0 & 0.000800 & 	$1.815726 \times 10 ^{-31}$ \\
\hline
LARES 2 &	0.002 & 12270000.0 & 0.002500 &	$1.882061 \times 10^{-33}$ \\
\hline
\end{tabular}
\end{table}

\section{Nutation}

In \cite{Shant}, it has been stated that besides the LT precession the nutation is the other important effect which can influence the orbital motion. Accordingly, the nutation is defined as

\begin{equation}\label{Nu}
\Omega_n = \frac{4 \pi}{T}
\end{equation}
where $T$ is the orbital period of Keplerian motion. Considering the $\Lambda$-gravity, the pure Keplerian dynamics is modified to
\begin{equation}\label{Kep}
T^2 = \frac{4 \pi^2}{\frac{GM}{r^3} - \frac{\Lambda c^2}{3}}
\end{equation}
Consequently, we can obtain error limits of $\Lambda$ from Keplerian dynamics as a weak-field limit effect
\begin{equation}\label{KepEr}
\Lambda\leq \frac{3}{c^2} (\frac{GM}{r^3} (1 \pm \frac{\Delta T}{T})^{-2} - \frac{GM}{r^3})
\end{equation}

The main advantage of the above relation is that we can use it to study several Keplerian systems i.e. from satellites orbiting around the Earth to S-Stars orbiting around Sgr A* \cite{DP}. In general, the vicinity of Sgr A* is considered as an important area for testing GR via various effects such as e.g. via pulsar timing  \cite{I1,I2}.

 In this work, we have used the recently updated data of exoplanets \cite{Exo} and computed the upper limits for $\Lambda$ which are tabulated in \ref{tab5}.

\begin{table}
\caption{}\label{tab5}
\centering
\begin{tabular}{ |p{1.8cm}||p{1.6cm}|p{1.6cm}|p{1.6cm}|p{3cm}|p{3cm}|p{2.4cm}|}
\hline
\multicolumn{7}{|c|}{Error limits of $\Lambda$ for Exoplanets} \\
\hline
Exoplanet & Mass of star ($M_{\odot}$) & semi major axis (AU) & 	Period (days) &	upper error of Period &	lower error of Period  &$\Lambda $ ($m^{-2}$)$<$   \\
\hline
DS Tuc A b & 0.959 & 0.0795 & 8.138268 & 0.00001 & -0.00001 & $6.229538 \times 10^{-33}$\\
\hline
GJ 758 b & 0.970 & 33 & 109000 & 198000 & -62000 & $1.569605 \times 10^{-34}$ \\
\hline
Kepler-538 b & 0.960 & 0.3548 & 81.73778 & 0.00013 & -0.00013 & $9.080541 \times 10^{-35}$  \\
\hline
kappa And b & 2.800 & 100 & 215000 & 100000 & -100000 & $9.279319 \times 10^{-36}$ \\
\hline
\end{tabular}
\end{table}

\begin{table}
\caption{}\label{tab6}
\centering
\begin{tabular}{ |p{1.8cm}||p{1.6cm}|p{1.6cm}|p{1.6cm}|p{3cm}|p{3cm}|p{2.4cm}|}
\hline
\multicolumn{6}{|c|}{Error limits of $\Lambda$ for S-Stars} \\
\hline
S-star  & semi major axis (AU) & 	Period (years) &	upper error of Period &	lower error of Period  &$\Lambda $ ($m^{-2}$)$<$   \\
\hline
S2  & 970.0 & 15.24 & 0.36 & -0.36 &  $3.050141 \times 10^{-34}$ \\
\hline
S6  & 5205.0 & 192.00 & 0.17 & -0.17 & $7.147862 \times 10^{-38}$  \\
\hline
S9  & 2156.0 & 51.30 & 0.70 & -0.70 & $1.580187 \times 10^{-35}$   \\
\hline
S12  & 2165.0 & 54.40 & 3.50 & -3.50 & $7.968392 \times 10^{-35}$  \\
\hline
\end{tabular}
\end{table}

We also obtain the upper limit of error for $\Lambda$ by studying the motions of S-Stars \cite{S-Star}. The results are shown in Table \ref{tab6} (we consider the mass of Sgr A* equal to $4.28 \times 10^{6} M_{\odot}$).

Finally, we check the error limits for $\Lambda$ at smaller scales. We analyze $\Lambda$-gravity corrections for satellites orbiting around the Earth. In this case, since there is no error limits for the orbital periods of satellites, we use the error limits of the Earth's mass i.e. 

\begin{equation}\label{Emass}
(5.9722 \pm 0.0006) \times 10^{24} \ \text{Kg}.
\end{equation}    
Considering the reported data of satellites analyzed in the previous section and also of the Moon, we get  the limits of $\Lambda$ as it is shown in Table \ref{tab7}.
\begin{table}
\caption{}\label{tab7}
\centering
\begin{tabular}{ |p{1.8cm}||p{3cm}|p{1.8cm}| p{2.4cm}|}
\hline
\multicolumn{4}{|c|}{Error limits of $\Lambda$ for objects orbitting the Earth} \\
\hline
Object&  semi major axis $(m)$ & Period  &$\Lambda $ ($m^{-2}$)$<$   \\
\hline
LAGEOS-1 &  12271150.0 & 13542.0 & 	$7.219382 \times 10^{-28}$  \\
\hline
LAGEOS-2 &	12161840.0 & 13347.6 & $7.415799 \times 10^{-28}$ \\
\hline
LARES &  7822000.0 & 6885.0 & 	$2.787420 \times 10^{-27}$ \\
\hline
LARES 2 &  12270000.0 & 13530.0 &	$7.221412 \times 10^{-28}$ \\
\hline
Moon & 384399000.0 & 2360592.0 & $2.348600 \times 10^{-32}$ \\
\hline
\end{tabular}
\end{table}

Moreover, we can extend the following analysis to the Solar System. Namely, in this case we have to use the error limits of the Sun's mass i.e. 
\begin{equation}\label{ESun}
(1.98847 \pm 0.00007) \times 10 ^ {30} \ \text{Kg}.
\end{equation}
Table \ref{tab8} illustrates the limits of $\Lambda$ in Solar System. By studying such cases, we show that it is possible to get constraints for $\Lambda$ which are even tighter than some relativistic effects. 
\begin{table}
\caption{}\label{tab8}
\centering
\begin{tabular}{ |p{1.8cm}||p{3cm}|p{2.4cm}|}
\hline
\multicolumn{3}{|c|}{Error limits of $\Lambda$ for objects in Solar System} \\
\hline
Planet& distance $(m)$ & $\Lambda $ ($m^{-2}$)$<$   \\
\hline
Mercury & $6.981708 \times 10^{10}$ & $4.573173 \times 10^{-34}$ \\
\hline
Venus  & $1.089418 \times 10^{11}$ &  $1.203700 \times 10^{-34}$ \\
\hline
Earth &	$1.520977 \times 10^{11}$ & $4.423181 \times 10^{-35}$ \\
\hline
Mars &	$2.492287 \times 10^{11}$ &	$1.005329 \times 10^{-35}$ \\
\hline
Jupiter & 	$8.160815 \times 10^{11}$ & $2.863532 \times 10^{-37}$ \\
\hline
Saturn & $1.503983 \times 10^{12} $ & $4.574818 \times 10^{-38}$ \\
\hline
Uranus & $3.006389 \times 10^{12} $ & $5.727524 \times 10^{-39}$ \\
\hline
Neptune & $4.536874 \times 10^{12}$ & $1.666604 \times 10^{-39}$ \\
\hline
Pluto 	& $7.375928 \times 10^{12}$ & $3.878400 \times 10^{-40}$ \\
\hline
\end{tabular}
\end{table}

\section{Conclusion}

In this paper we have analyzed the corrections of $\Lambda$-gravity for the orbital LT precession and the nutation. We have obtained the upper error limits of $\Lambda$ for both effects. Considering the LT effect, we have studied the data of four satellites aimed to measure the LT effect i.e. LAGEOS-1, LAGEOS-2, LARES and forthcoming LARES 2. This analysis have been performed for two different cases i.e. for a simplified toy model as well as for more realistic model. Interestingly, we have found that for the latter we can get an error limit of $\approx 10^{-33}$. This bound is stronger than upper bounds obtained by several other effects such as gravitational redshift, gravitational time delay and geodetic precession in Solar System \cite{L}. In the context of $\Lambda$-gravity we have obtained the constraints for $\Lambda$ in non-relativistic (weak-field) regimes which enables us to study the corrections of $\Lambda$ term in several astrophysical scales. Namely, by considering $\Lambda$-gravity we have studied the Keplerian systems from satellites orbiting around the Earth to S-Stars which are moving around the central black hole of Milky Way i.e. the Sgr A*. For all of them, the obtained constraints are in agreement with the numerical value of $\Lambda$ obtained by cosmological observations. Since modified gravity theories can predict certain effects differently, the testing of the validity of each effect to observations, also at various available scales, has no alternatives.

\section*{Acknowledgement}
We are thankful to the referee for insightful comments and suggestions.

\end{document}